\documentclass[epjCONF]{svjour}
\usepackage{graphics}
\usepackage[varg]{txfonts} 
\usepackage[latin1]{inputenc}
\session-title{Conference Title, to be filled}
\begin{document}
\title{Constraints to dark-matter properties from asteroseismic analysis of KIC 2009504\footnote{Conference proceeding to be published in 
EPJ Web of Conferences. The corresponding article is in press to Physical Review D, arXiv:1505.01362.}}
\author{Brand\~ao, I., M.\inst{1}\fnmsep\thanks{\email{isa@astro.up.pt}} \and Casanellas, J.\inst{2}}
\institute{Instituto de Astrof\'isica e Ci\^encias do Espa\c{c}o, Universidade do Porto, CAUP, Rua das Estrelas, 4150-762 Porto,
Portugal \and Max Planck Institute for Gravitational Physics, Albert Einstein Institute, Am M\"uhlenberg 1, 14476, Golm, Germany}
\abstract{Asteroseismology can be used to constrain some properties of 
dark-matter (DM) particles \cite{casanellas13}. In this work, 
we performed an asteroseismic modelling of the main-sequence solar-like 
pulsator KIC 2009505 (also known as Dushera) in order
to test the existence of DM particles with the characteristics that
explain the recent results found in some of the DM direct detection 
experiments. We found that the presence of a convective core in KIC 2009504
is incompatible with the existence of some particular models of DM particles.} 
\maketitle
\section{Introduction}
\label{intro}
Stars slightly more massive than the Sun ($M > 1.1 \rm M_\odot$) develop a tiny convective core during the
main sequence. In case of DM particles do not self-annihilate inside stars their accumulation 
leads to an efficient mechanism of energy transport that results in the suppression of
the convective core expected to be present in main-sequence stellar models with masses between 
1.1 and 1.3 $\rm M_\odot$ in a dark-matter free scenario \cite{casanellas13}.

The presence of a convective core leaves a signature on the oscillation frequencies of a pulsating
star, which, in principle, can be detected through the use of asteroseismic diagnostic tools, such as 
$dr_{0213}$, $r_ {010}$ and $r_{02}$ (e.g., \cite{cunha07,cunha11,brandao14}).
Their frequency derivative (the \lq slope\rq) is expected to provide information about convective core's 
properties and stellar age \cite{brandao14}.  

In this work, we test the existence of DM models with a given physics that may explain the positive 
results of the direct detections in the DAMA and CoGeNT experiments \cite{buckley13}.
\section{Method}                              
\label{sec:1}
To model KIC 2009505, we computed three grids of main-sequence evolutionary tracks using the
CESAM code \cite{morel08}. Grid 1 considers a free DM scenario while grids 2 and 3 consider 
two different DM scenarios following the prescription of \cite{gould87} 
for the capture rate and of \cite{gould90} 
for the energy transport by DM conduction. We considered a mass range of [1.1,3.1] $\rm M_\odot$
(0.005 steps), an abundance of heavy elements range of [0.012,0.024] (0.0005 steps),
we fixed the overshooting parameter to 0.1 and the mixing-length parameter \cite{bohm-vitense58} to 1.8.  
The physics used in the code was the same as described in \cite{casanellas13}.
Diffusion was taken into account following the prescription by \cite{michaud93}. 
The model oscillation frequencies were computed using the Aarhus adiabatic 
oscillation code ADIPLS \cite{cd08}.

As non-seismic constraints to the modelling we considered the effective temperature, $T_{\rm eff}$, to be $6200\pm200$, 
the logarithm of gravity, $\log g$, to be $4.30\pm0.2$, and 
the initial abundance of heavy elements to hydrogen ration, $Z/X_{\rm s}$, to be $0.023\pm0.09$. 
For seismic constraints we considered the mean of the large frequency separation (e.g., \cite{roxburgh03}) for degrees 
$l = 0-2$ computed in the range of the observed frequencies, $<\Delta\nu>_{012}$, as $88\pm0.6$ and the absolute
value of the slope of the diagnostic tool $\Delta\nu_0 r_{010}$, $|S\{\Delta\nu_0 r_{010}\}|$, 
as $0.0032\pm0.0006$ (see \cite{brandao14} for details). Both non-seismic and seismic constraints
were taken from \cite{silva-aguirre}.
\section{Results}
\label{sec:2}
We found that all models in a DM free scenario show a convective core, 
with a mean radius of $0.06 \pm 0.01 \rm R_\star$ in agreement with \cite{silva-aguirre}. 
Similar results were observed in DM models whose particles have the properties that explain 
the positive results in the DAMA experiment. However,
when the  influence of the existence of DM particles with the properties that explain
the positive results in the CoGeNT experiment was taken into account, 
none of the models of KIC 2009505 had a convective core. Moreover,
we found that the average value of $|S\{\Delta\nu_0 r_{010}\}|$  computed for these set of DM models
is in disagreement with observations as these models do not show a convective core.
\section{Conclusions}
\label{sec:3}
The existence of asymmetric DM particles with the properties that explain the positive results in the
CoGeNT experiment would lead to the suppression of the convective core recently detected in the
main-sequence solar-like pulsator KIC 2009505. We have shown that the sensitivity of the slope of
the diagnostic tool $\Delta\nu_0 r_{010}$  to the presence of a convective core can be used to rule out the existence
of such DM models even when these models reproduce well the observed large frequency separation.
\begin{acknowledgement}
I.M.B. acknowledge the support of the Funda\c{c}\~ao para a Ci\^encia e a Tecnologia (FCT) in the form
of the grant reference SFRH/BPD/87857/2012 and acknowledges the support from the EC 
Project SPACEINN (FP7-SPACE-2012-312844). J.C. acknowledges the support from the Alexander von Humboldt Foundation.
\end{acknowledgement}
\end{document}